\documentclass[aps,draft,superscriptaddress,twocolumn,showpacs]{revtex4}

\usepackage{bm,graphicx,amsmath,amssymb}
\usepackage{textcomp}

\def \be{\begin{equation}}
\def \ee{\end{equation}}

\DeclareGraphicsExtensions{.jpeg,.pdf,.tif,.gif}

\begin{document}

\preprint{}

\title{Giant Anisotropic Magneto-Resistance in ferromagnetic atomic contacts}
\author{M. Viret}
\affiliation{Service de Physique de l'Etat Condens{\'e}, CEA
Saclay, F-91191 Gif-Sur-Yvette}
\author{M. Gabureac}
\affiliation{Service de Physique de l'Etat Condens{\'e}, CEA
Saclay, F-91191 Gif-Sur-Yvette}
\author{F. Ott}
\affiliation{Laboratoire Leon Brillouin, CEA Saclay, F-91191
Gif-Sur-Yvette}
\author{C. Fermon}
\affiliation{Service de Physique de l'Etat Condens{\'e}, CEA
Saclay, F-91191 Gif-Sur-Yvette}
\author{C. Barreteau}
\affiliation{SPCSI, CEA Saclay, F-91191 Gif-Sur-Yvette}
\author{R. Guirado-Lopez}
\affiliation{Instituto de Fisica, Universidad Autonoma de San
Luis Potosi, Alvaro Obregon 64 78000 San Luis Potosi, Mexico}

\begin{abstract}

Magneto-resistance is a physical effect of great fundamental and industrial interest since it is the basis for the magnetic field sensors used in computer read-heads and Magnetic Random Access Memories. As device dimensions are reduced, some important physical length scales for magnetism and electrical transport will soon be attained. Ultimately, there is a strong need to know if the physical phenomena responsible for magneto-resistance still hold at the atomic scale. Here, we show that the anisotropy of magneto-resistance is greatly enhanced in atomic size constrictions. We explain this physical effect by a change in the electronic density of states in the junction when the magnetisation is rotated, as supported by our ab-initio calculations. This stems from the "spin-orbit coupling" mechanism linking the shape of the orbitals with the spin direction. This sensitively affects the conductance of atomic contacts which is determined by the overlap of the valence orbitals.

\end{abstract}

\pacs{PACS numbers: 75.70.Kw, 72.75.Gd}

\maketitle

The effect of an external field on the resistivity of pure
ferromagnetic metals (the magnetoresistance-MR) was the subject
of intense research work in the second half of the 20ieth
century. The field has seen a renewed interest in the past
fifteen years with the discovery of giant effects in systems
combining magnetic and non-magnetic materials.
This Giant Magneto-Resistance (GMR) has had a tremendous impact
both through its industrial applications as read-heads and
Magnetic Random Access Memories as well as for triggering the field of "spintronics"
\cite{barthelemy02}, aiming to use the spin of the charge carriers in
electronic devices with higher functionalities. As the pressure
towards miniaturization increases, it is important to understand
how magnetoresistive effects are influenced by size reduction. In
constrictions of dimensions close to the Fermi wavelength,
boundary conditions enforce that transverse electronic modes are
quantized which results in the discreteness of propagating
electron modes. 2-D electron gases are archetypical systems in
which the conductance is quantized in units of $2e^{2}/h$. In
metals where the Fermi wavelength is typically 3 \AA, one needs
to reach atomic dimensions in order to observe such effects
\cite{agrait03}. But because even in the single atomic regime
several orbitals overlap, one normally finds that several conduction
channels are opened with imperfect transparency, i.e. each channel has a transmittance associated to it (a coefficient between 0 and 1). Calculations seem to indicate that
4 or 5 channels participate significantly to the conduction of 3d
transition metal nanocontacts \cite{smogunov03,bagrets04}. The
magneto-resistance obtained when one side of the contact flips
its magnetization is enhanced compared to that in the bulk
\cite{bagrets04,palacios04} and values of the order of 20\% have been
reported in some careful experiments \cite{viret02,yang04}. On
the other hand, one could expect some dependence of the
conductance to the direction of the magnetization because
changing the spins' direction will affect the orbitals through a
mechanism known as spin-orbit coupling. One can describe this
interaction by a term in the system's Hamiltonian written
$\lambda{\vec L}.{\vec S}$ where $\lambda$ is called the
spin-orbit constant, ${\vec L}$ and ${\vec S}$ are the orbital
momentum and spin vectors. In a solid, molecular orbitals lose
most of their angular momentum, an effect known as "quenching".
As a result, the resistance variation with the angle between the
local moments and the electrical current lines: the Anisotropic
Magneto Resistance (AMR), is only at most a few percent. This is
actually the oldest known magnetic effect on electronic transport
in ferromagnets,which was discovered in 1857 by W. Thomson
\cite{ThomsonAMR}. When the dimensionality of the system is reduced, like in mono-atomic
wires, orbital moments are much larger than in the bulk
\cite{gambardella02}. One can then expect spin-orbit mechanisms,
like the AMR, to be enhanced. Unfortunately, it is extremely
difficult to study electrical transport in these systems because
one needs to contact tiny structures, a task that often turns out
to be impossible.

\begin{figure}[ht]
\includegraphics*[scale=0.35,draft=false,clip=true]{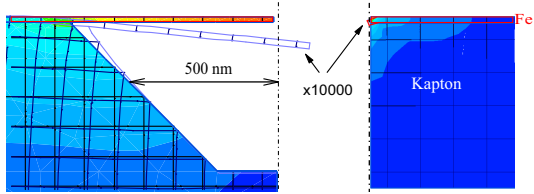}
\caption{Magnetostriction induced deformation of the nano-bridges in two geometries: (left) suspended and (right) attached to the kapton substrate (only half of the structures is represented - the other half is obtained by symmetry). The grey/colour scale represents the deformation which reaches $15 pm$ for the suspended bridge and $1 pm$ when not under-etched (the field is along the bridge). For visibility, the structure with distortions magnified by 10000 is also shown.} \label{figdist}
\end{figure}

Among the techniques suitable for studying electronic transport
in reduced dimensions, break junctions represent an interesting
option, because they provide an easy way to drive a current
through only a few atoms. The measurement methods we used are
based on the breaking of a nanoscopic structure in a controlled
manner while monitoring its resistance \cite{vanruitenbeek96}. The setup is particularly stable since most of the structure is attached to the substrate
and only the narrow bridge to be broken is suspended. It is then
possible to mechanically stabilize the contact with a precision
of a few picometres. We have slightly modified this procedure and
decided not to suspend the bridge, because in ferromagnetic
materials, any unsupported part is subjected to a magnetically
induced distortion called magnetostriction. This effect results
in a modification of the contact geometry when the magnetization
changes direction, which has been shown to significantly affect
the contact resistance \cite{Gabureac04}. This difference in the
bridge fabrication is essential here as it proves to be very
efficient to reduce magneto-mechanical effects. Fig. \ref{figdist}
shows the result of a finite elements numerical simulation where
a 20nm Fe layer is deposited on a kapton substrate with and
without the under-etching procedure. Taking the physical
parameters for the two materials and a saturation
magnetostriction of polycrystalline iron of $-8 ppm$, the
distortion of the Fe layer at the contact level goes from $15 pm$
when the bridge is suspended to $1 pm$ when the structure is
attached to the kapton (no under-etching). We have checked that
even in the tunnelling regime (where R changes exponentially with
the gap), the effect is negligible. Direct forces due to stray fields are also found to be two orders of magnitude smaller. In fact, any variation in the temperature of the electrodes would have a greater consequence than the magneto-mechanical effects. Therefore, the procedure insures that, in applying a saturating rotating field, the resistance change is only due to the intrinsic anisotropic "AMR" effect.

We have carried out an extensive set of measurements on Fe break
junctions (similar results are also obtained for Ni and Co),
where a 2.5 T magnetic field is rotated in the plane of the
contact while the resistance is monitored. Interestingly, the
resistance and the amplitude of its angular change did not
significantly depend on the field magnitude between 0.5 and 2.5 T,
where magnetostriction is expected to change by about 50\%. This
provides further experimental evidence that the effect can be
neglected. In order to measure changes of resistance with
magnetization direction, a high field was chosen to make sure the
atomic contact is always in its saturated state (the demagnetization can be large in nanostructures).

\begin{figure}[ht]
\includegraphics*[scale=0.35,draft=false,clip=true]{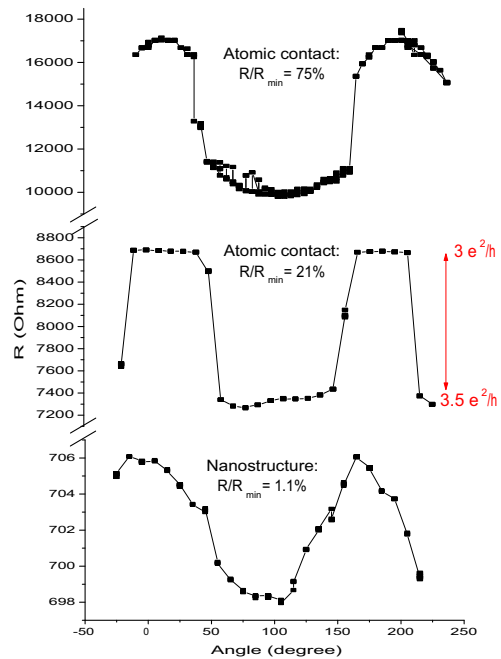}
\caption{Variation of resistance as a 2.5 T field is rotated in
the plane of the contact. The bottom graph is obtained in the
first stages of pulling the bridge (as the nanostructure is not
yet broken). There, the AMR is close to the bulk value, i.e.
around 1\%. In the atomic contact regime, at a conductance of
$3e^{2}/h$ in the middle graph the AMR behaviour is close to a two
level effect reaching $3.5e^{2}/h$ when the magnetization is
perpendicular to the contact. In the top graph, where the
conductance is close to $2e^{2}/h$, the effect is intermediate
between the two behaviours and reaches 75\%} \label{fig1}
\end{figure}

Figure \ref{fig1} shows a representative set of curves in Fe at 4.2 K. Interestingly, a behaviour qualitatively different from the $cos^{2}(\theta)$ dependence of the bulk, can be observed in the atomic contact regime. In the middle graph, it looks likely that one channel gets blocked when the field is along the contact, leading to a two-level conductance and an atomic-AMR (AAMR) effect of 21\%. At slightly different values of conductance, both smooth sinusoidal variations as well as discrete jumps are observed (see top graph). This is to be expected when overlap changes are not sufficient to completely close a channel, but enough to change their transmittance. This general behaviour is consistent with what is known to happen for non magnetic break junctions when orbital overlap is varied by mechanical deformation of the contacts \cite{Scheer98}. There, theoretical calculations have shown that both sharp jumps and smooth variations of the conductance can be explained by considering the details of orbitals overlap \cite{Cuevas98}.

\begin{figure}[ht]
\includegraphics*[scale=0.38,draft=false,clip=true]{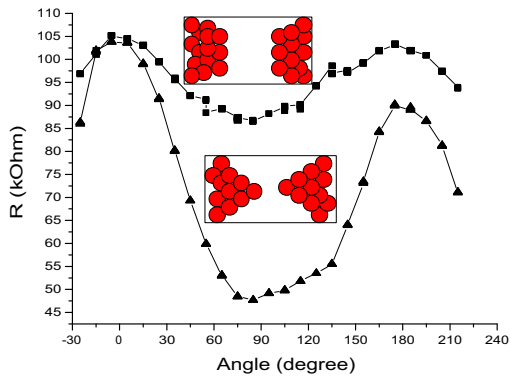}
\caption{Measured resistance variation in Fe atomic contacts in the tunnelling regime when a 2.5 T field is rotated in-plane. The large 100\% AMR effect is obtained at a shorter gap value than that for the 20\% effect of similar resistance.} \label{figtunnel}
\end{figure}

Even more surprising, is the effect measured in tunnelling and shown in figure \ref{figtunnel}. In this regime, charge carriers jump from one electrode to another through a very narrow vacuum gap (a few \AA at most in break junctions). There, the evanescent wave functions still have a strong atomic orbital character from which they can inherit the spin-orbit coupling properties. Moreover, because of the exponential decrease of the wave intensity with distance and the corresponding exponential increase of resistance, it is likely that any change in the shape of the orbitals could have a more significant effect. This is indeed what we experimentally observe. Interestingly, the two measurements shown in Fig.\ref{figtunnel} correspond to electrons tunnelling through gaps of different sizes (as measured in our setup) and similar resistance. Hence, the tunnelling cross section must differ which implies that electrons tunnel from sharp tips and small gap or flat tips and large gap. The atomic orbital character of the evanescent waves is obviously stronger in the first case, and one can expect the tunnelling-AMR (TAMR) to be larger, as measured. Moreover, because evanescent d orbitals are generally shorter range than s orbitals their contribution should be larger for narrow gaps.

\begin{figure}[ht]
\includegraphics*[scale=0.4,draft=false,clip=true]{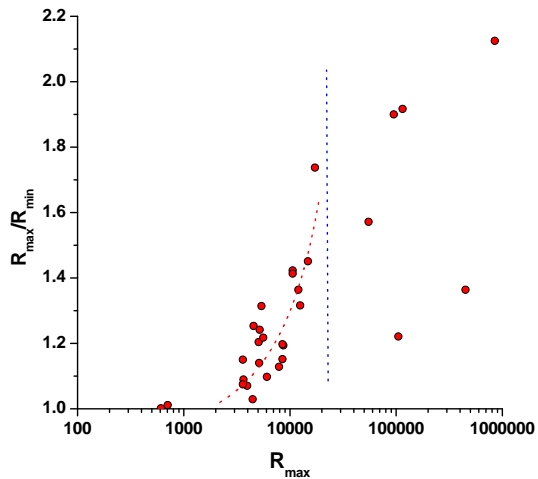}
\caption{Amplitude of the AMR effect in contacts of different resistances. The dotted line is a guide to the eye underlining the steep increase as the contact reaches atomic sizes. The vertical
line is at $e^{2}/h$ and represents a rough separation between tunnelling and atomic contact regimes.} \label{fig3}
\end{figure}

The measured ratio of high to low resistances (when rotating the field) for a Fe sample in different atomic configurations are gathered in figure \ref{fig3}. This plot is instructive because it shows that the AMR effect increases steeply as the contact reaches atomic dimensions. For higher resistances (above $e^{2}/h$), when in the tunnelling regime, we observe a significant scatter in the amplitude of the TAMR where values around 100\% can be achieved but effects as low as 20\% can also be found. As explained above, this can be understood because the atomic orbital character of the evanescent waves depends on the exact atomic configuration, which can vary appreciably at constant resistance.

Theoretically, a considerable amount of work has been devoted to metallic atomic contacts, but magnetism has seldom been considered. Most relevant works have studied the resistance generated by a "magnetic domain wall" on the contact \cite{smogunov03} and very recently, the contribution from the exchange splitting has been found to be rather small \cite{bagrets04}, the dominating effect being instead the orbital nature of the conduction electrons \cite{palacios04}. In order to understand the origin of AAMR effect, we have performed ab-initio calculations to determine the changes of the electronic band structures with the spins direction. In an effort to give a
pedagogical illustration of the phenomenon, we consider an ideal atomic chain of Fe atoms. The electronic structure of the wire is obtained using the pseudo-potential plane-wave method implemented in the PWscf package \cite{PWSCF} which, in its last version, allows to include spin-orbit interactions. Due to the use of plane waves, the system considered is in fact a periodic array of atomic chains, for which the distance between two wires is large enough (15 \AA) to avoid interactions. We have first carried out ultrasoft pseudo-potential calculations without spin-orbit in the Generalized Gradient Approximation (GGA) to obtain an equilibrium
spacing $a$ of the atoms in the wire of $2.27$ \AA and a magnetic moment of 3.3 $\mu_B$ per atom. Then using a fully relativistic ({\sl i.e} solving the Dirac equation for an atom) ultrasoft pseudo-potential including spin-orbit coupling \cite{dalcorso05} in the Local Density Approximation (LDA) we have calculated the electronic structure of the wires with a lattice spacing of $a$.
\begin{figure}
\includegraphics*[scale=0.43,angle=0,draft=false,clip=true]{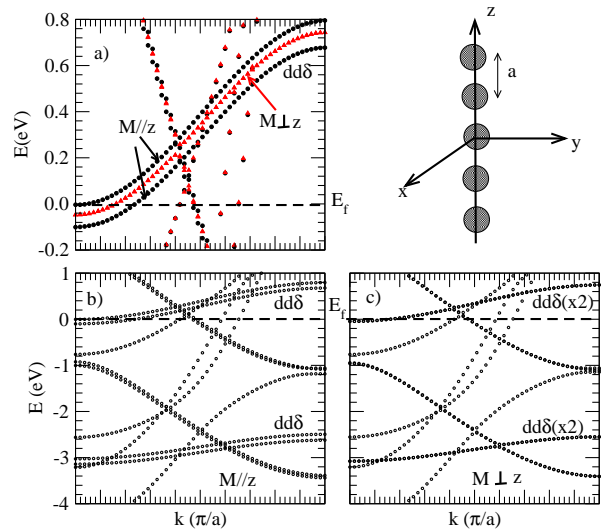}
\caption{Calculated band structure for a Fe mono-atomic chain (see
top right inset) magnetized in the parallel b) and  perpendicular
c) directions. In a) the band structure for a parallel (full
black circles)  and perpendicular magnetization (red empty
triangles) is shown at a larger scale around the Fermi level. }
\label{fig4}
\end{figure}
Fig. \ref{fig4} shows the band structure obtained for the
magnetization parallel and perpendicular to the chain.
Significant changes near the Fermi level are observed, which stem
from the degeneracies induced by spin-orbit coupling. The
interesting bands are the weakly dispersive, so called Slater
Koster $dd\delta$ bands, which have a pure $d_{xy}$ (and
$d_{x^2-y^2}$) character since they couple neither to $s$ and $p$
nor to other $d$ states with different symmetry. It is clear from
Fig. \ref{fig4} that these bands split by about 0.10 to 0.12 eV
when the magnetization is rotated from perpendicular to parallel
to the atomic wire. It can actually be shown in a tight-binding
model \cite{MCD_DS_TB} that the splitting is equal to $2
\lambda$, which leads to a value of $\lambda=50-60 meV$. Because
these states cross the Fermi level, they will play an important
role in defining conduction channels with some transmittance. Since
they are also very sensitive to the magnetic orientation, a
magnetoresistance effect can be expected. Indeed Fig. \ref{fig4}
shows that when the magnetization is perpendicular to the chain
the two degenerate $dd\delta$ bands are crossing the Fermi level
($E_f$) while in the parallel case their splitting pushes the
upper band to higher energies where it hardly touches $E_f$. This
results in a significant electronic transfer which almost empties
the upper band. Therefore we can expect a better conductivity
when the magnetization is perpendicular to the chain than when it
is parallel, in good agreement with the experimental findings. Of
course, a direct inference of the conductance from the band
structure topology is not possible, but these quantities are
intimately linked. What can then safely be said here is that the
transmission of some conduction channels will surely change with
the magnetization direction. Beyond this simple atomic chain
model, we have carried out preliminary tight-binding calculations
in a more complex geometry of two pyramids in contact on their
apex. The result is that the local density of states at the atoms
forming the contact region is also modified when changing the
orientation of the magnetization, although a bit less
significantly than for the 1-D wire. From these examples, we note
that the change in the energy level distribution is complex which
could induce a positive or negative AAMR effect, rather sensitive
to the atomic, magnetic, and electronic structure at the
constriction region. Hence, one can expect the AAMR to depend on
the exact atomic geometry as well as the potential difference
across the contact. One can even envision sign changes in the AMR
with applied voltage.

In conclusion, we have shown that rotating a saturating field in
an iron atomic contact leads to significant resistance changes.
The effect is due to a spin-orbit coupling induced modification of
orbitals overlap. This is supported by ab-initio calculations
showing clear changes of the electronic band structure when the
orientation of the magnetization in the samples is rotated by
90 \textdegree. The AMR effect also exists in the tunnelling regime where
evanescent orbitals keep an atomic character. The measured
amplitude of the effect opens the possibility of using it in
field sensors or magnetic memories of atomic size. We find that
the AMR effects (both AAMR and TAMR) are actually larger than the
spin scattering effects obtained when the magnetizations are
opposite on the two sides of the atomic contact.

\acknowledgments

It is our pleasure to thank M.C. Desjonqu\`eres and D. Spanjaard
for fruitful discussions on the tight-binding model and A. Dal
Corso for providing a fully relativistic ultrasoft pseudo
potential including spin-orbit coupling for Iron. We would also
like to thank O. Klein for his experimental help and valuable
discussions and X. Waintal for very interesting theoretical
conversations.

\end{document}